\documentclass[aps,prl,twocolumn,groupedaddress,10pt]{revtex4-1}
\usepackage{amsmath}
\usepackage{graphicx}
\usepackage{float}
\usepackage{epstopdf}
\begin{document}
\title{Is the pseudogap a topological state?}
\author{Alfredo. A. Vargas-Paredes}
\affiliation{Departamento de F\'{\i}sica dos S\'{o}lidos, Universidade Federal do Rio de Janeiro, 21941-972 Rio de Janeiro, Brazil}
\author{Marco Cariglia}
\affiliation{Departamento de F\'{\i}sica , Universidade Federal de
Ouro Preto, 35400-000 Ouro Preto Minas Gerais, Brazil}
\author{Mauro M. Doria}
\affiliation{Departamento de F\'{\i}sica dos S\'{o}lidos, Universidade Federal do Rio de Janeiro, 21941-972 Rio de Janeiro, Brazil}
\date{\today}
\begin{abstract}
We conjecture that the pseudogap is an inhomogeneous condensate above the homogeneous state whose existence is granted by topological stability. We consider the simplest possible order parameter theory that provides this interpretation of the pseudogap and study its angular momentum states. The normal state gap density, the breaking of the time reversal symmetry and the checkerboard pattern are naturally explained under this view. The pseudogap is a lattice of skyrmions and the inner weak local magnetic field falls below the experimental threshold of observation given by NMR/NQR and  $\mu$SR experiments.
\end{abstract}


\maketitle

\section{Introduction}
The discovery of the high temperature (high $T_c$) superconductors by Bednorz and M\"uller (1986)~\cite{bedmull} brought new paradigms to the field of condensed matter physics. Three-dimensional superconductivity  originates in two-dimensional layers where Cooper pairs are formed and outlive elsewhere. The understanding of any layered compound demands its study at several doping, achieved by changing the number of carriers available in the layers. These properties are best displayed in the so-called temperature versus doping diagram where the critical temperature $T_c$ defines a dome shaped curve whose onset and disappearance takes place at critical doping values. In this phase diagram superconductivity is one among many other possible electronic states that involve magnetic, charge and pairing  degrees of freedom that either coexist, cooperate or dispute the same spatial locations within the layers.
Besides the superconducting and the anti ferromagnetic state there is another characterized electronic state, the pseudogap state, that lives below  a temperature $T^*$ versus doping line in this phase diagram. Interestingly this line approaches the superconducting dome from above by increasing the doping, always with a negative slope and, at some doping value, intersects and crosses the dome, ending at zero temperature, where it defines a so-called quantum critical point.
The pseudogap was revealed in 1989, soon after the discovery of Bednorz and M\" uller, by the observation of a sharp decrease of the nuclear spin susceptibility in the cuprate layer atoms (CuO$_2$)~\cite{alloul89}.
This  sharp decrease of the NMR Knight shift K indicated the existence of
a (above $T_c$) normal state gap, the pseudogap.
The nature of the pseudogap remains so far unknown and a true challenge to the field. It is not clear whether the pseudogap gap line $T^*$ is a thermodynamical or crossover transition~\cite{arkady13}. Among the known properties of the pseudogap, are the normal state gap and the spontaneously broken symmetries. The pseudogap state breaks the time reversal symmetry because  left-circularly polarized photons give a different photocurrent from right-circularly polarized photons below the line $T^*$, as shown more than ten years ago by Kaminski et al.~\cite{kaminski02}. Recently these results have been confirmed through high precision measurements of polar Kerr effect~\cite{xia08,he11}, which are also clearly suggestive of a phase transition at $T^*$, below which arises a finite Kerr rotation~\cite{hovnatan12}.
The pseudogap also breaks translational invariance symmetry within the layers and this modulation was first observed through scanning tunneling microscopy, initially coined as the checkerboard pattern, a tetragonal lattice with $4a$ periodicity, where $a$ is the CuO$_2$ unit cell length. This periodicity was  firstly found inside the vortex cores~\cite{hoffman02} of Bi$_2$Sr$_2$CaCu$_2$O$_{8+x}$, but soon after incommensurate patterns were also observed in the normal state of this compound~\cite{vershinin04}, below the pseudogap line in the absence of a magnetic field. It is quite remarkable that in the pseudogap phase carriers within the layers arrange themselves into a periodic pattern not necessarily commensurate with the crystallographic structure~\cite{mcelroy05}.
Nowadays the checkerboard pattern is seen as a consequence  a charge density wave~\cite{li06}.
Recent work using resonant elastic x-ray scattering (REXS) correlation done by  G. Ghiringhelli et al.~\cite{ghiringhelli12} found concrete evidence of this charge-density-wave in the underdoped compound
YBa$_2$Cu$_3$O$_{6+x}$ with an incommensurate periodicity of nearly $3.2a$, both above and below $T_c$.
Thus it is quite clear that a theoretical attempt to explain the pseudogap must take into account the normal state gap and also the broken symmetries.
It happens that a magnetic order is expected to arise due to the breaking of the time reversal symmetry. Time reversal symmetry means that a state remains unchanged upon time reversal. We know that the velocity (momentum) and also the magnetic field both reverse sign under a time-reversal operation. Magnetic spins reverse direction when time reverses direction, therefore magnetic order breaks time reversal symmetry and may be the reason for it. The presence of any kind of magnetic order leads to a local magnetic field that must be experimentally accessed by several magnetic probes. Indeed there has been an intense search for this predicted spontaneous local magnetic field inside the high $T_c$ superconductors in the pseudogap phase. Polarized neutron diffraction experiments~\cite{fauque06,li11} indicate a magnetic order below the pseudogap. NMR/NQR~\cite{strassle08,strassle11} and  $\mu$SR~\cite{macdougall08,sonier09} experiments set an upper limit to the
magnetic field between the layers, which must be smaller than $0.1 \; \mbox{Gauss}$. A proposal of magnetic order with the layers has been put forward by C. M. Varma~\cite{varma06,fauque06,li11,bourges11}, who claims that microscopic orbital currents cause this order, and consequently, the breaking of the time reversal symmetry.

Recently we have shown that an above the homogeneous (normal) state gap and the broken symmetries can be understood in the context of an order parameter description, and for this reason, suggested this interpretation for the pseudogap~\cite{cariglia14}. Thus the pseudogap is a gapped topological state made of skyrmions, whose microscopic nature is still controversial. We recall that the description of a condensate through an order parameter was proposed by Ginzburg and Landau much before the BCS unveiled the microscopic mechanism behind superconductivity. According to Gorkov and Volovik~\cite{volovik85} a system described by an order parameter that breaks the time reversal symmetry must have an accompanying  magnetic order that yields  a local magnetic field near to the sample surface even in the absence of an external field. Indeed our description of the pseudogap  exactly fits the Gorkov and Volovik scenario with the local magnetic field  found around the layers and not just near to the sample surface. This magnetic field originates from spontaneous circulating supercurrents in the layers that give rise to an inhomogeneous excited state which is stable since it is prevented from  decaying into the homogeneous ground state by its topological stability. This tetragonal lattice of skyrmions  breaks time reversal symmetry and also  translational invariance, has an energy gap above the homogeneous state that we associate to the pseudogap. We obtain the numerical value of the pseudogap density as a function of the local magnetic field between the layers, which is assumed to fall below the experimental threshold of observation set by NMR/NQR~\cite{strassle08,strassle11} and  $\mu$SR~\cite{macdougall08,sonier09} experiments. In this paper we provide a detailed study of this order parameter approach to the pseudogap state and derive its angular momentum properties.

\section{The theory}
We seek here the simplest possible theory able to describe a condensate, the pseudogap, through an order parameter such that it lies above the homogeneous state separated by a gap, hereafter called the normal state gap, and presents broken symmetries. Besides the supercurrents created by its inhomogeniety must be in conformity with the experimental threshold imposed by the maximum observed internal field.
Because of its assumed simplicity the present theory does not describe the transition at $T^*$, as we only try to capture the major qualitative features of the pseudogap. Thus the present description is restricted to temperatures below but near to the $T^*$ line, and above $T_c$, such that the condensate energy, which regulates the superconducting dome and defines $T_c$, can be safely ignored. Thermal fluctuations are not included here and so the discussion is restricted to mean field considerations.
Even in this simplified context we find that to describe both the pseudogap and the superconducting states there must be at least two components, $\Psi=\left(\begin{array}[c]{cc}\psi_u \\ \psi_d \end{array}\right)$. This simplest possible theory is just the sum of the kinetic and the field density energies,
\begin{equation}
F=\int\frac{d^{3}x}{V}\left [\frac{|\vec{D}\Psi|^2}{2m}+\frac{\vec{h}^2} {{8\pi}}\right ],
\label{f0}
\end{equation}
There is minimal coupling to the field through the covariant derivative $\vec D = (\hbar/i)\vec \nabla - (q/c)\vec A$, $\vec h = \vec \nabla
\times \vec A$, and $m$ and $q$ are the Copper pair mass and charge, respectively. Undoubtedly the lowest free energy state of Eq.(\ref{f0}) is the null homogeneous state $\Psi=0$, thus  with no supercurrents and no resulting local magnetic field, $\vec h=0$ that we identify with the superconducting state restricted to exist under the superconducting dome. Our claim is that above this homogeneous null state lives an inhomogeneous gapped state, topologically stable, conjectured to be the pseudogap. In this paper we give a detailed derivation of the pseudogap from Eq.(\ref{f0}) and obtain its angular momentum properties.

This simple theory has a global $SU(2)$ rotational invariance~\cite{doria10,doria12} that arises as a natural consequence of the presence of  two-components. We believe that this invariance is explicitly broken by extra terms in the free energy, not considered in this simple theory. Even  without considering this explicit breaking we find that this naively introduced $SU(2)$ symmetry has possibly only local meaning and is associated to the group of spatial rotations, which turns $\Psi$ into a truly spinorial order parameter. It is not generally possible to bring the two-component order parameter into a one-component form in all space, $\Psi'=U\Psi$, $\Psi'^T = \big ( \psi' \quad 0 \big)$ by a constant SU(2) rotation,  $UU^{\dagger}=1$. It is only possible to get rid of one of the component in case of an homogeneous solution. For a spatially inhomogeneous  solution, $\Psi(\vec x)$, the rotation to one component form can only be done locally, $U(\vec x)$. Therefore the free energy of Eq.(\ref{f0}) cannot be reduced to one component in case of an inhomogeneous solution. Under a local rotation $U(\vec x)$ an extra term appears in the covariant derivative, $\vec D'= \vec D + U\vec D U^{-1}$ so that $F_k = \int d^{3}x \left ( \vert \vec{D'}\Psi'\vert^2/2m \right )/V$. We consider this a signal of a non-abelian gauge symmetry in the pseudogap phase. This possibility has been considered elsewhere~\cite{alfredo13,marchetti11} but will not be treated here. From the free energy of Eq.(\ref{f0}) we obtain the following variational equations:
\begin{eqnarray}
\frac{\vec{D}^2\Psi}{2m}=&&0, \label{gl}\\
\vec{\nabla}\times\vec{h}=&&\frac{4\pi}{c}\vec{J}, \label{amp}
\end{eqnarray}
where the supercurrent density is given by
\begin{equation}
\vec{J}=\frac{q}{m}\left[\Psi^{\dag}\vec{D}\Psi+\left(\vec{D}\Psi\right)^{\dag}\Psi\right].
\end{equation}
The aforementioned weakness of the local magnetic field  is a key ingredient for the construction of a perturbation scheme to solve the above equations. The local magnetic field itself works as the small parameter helpful to solve these equations recursively by the following procedure. The order parameter $\Psi$ is obtained from Eq.(\ref{gl}) neglecting the interaction with the field, which means that the equation to solve is truly $\vec \nabla^2\Psi=0$. Once in power of this solution $\Psi$, and the supercurrent obtained in this order, Eq.(\ref{amp}) is used to derive $\vec h$ from a known source. This iterative procedure is repeated recursively, and so, defines a perturbative scheme. We carry it just to the first iteration, which means that the local magnetic field must be very weak. Another important ingredient to added in our search of the solution of the above equations is the geometry, namely, a stack of layers, such that the order parameter is assumed to evanesce away from each one of them. Thus we seek a Fourier solution of $\vec \nabla^2\Psi=0$, $\Psi=e^{i(\vec k\cdot \vec x+k_3 x_3)}\Psi_{(\vec k,k_3)}$, where the wave vector along the layer is,
\begin{equation}
\vec k \equiv k_1 \hat x_1 + k_2 \hat x_2, \mbox{and} \; k\equiv \vert\vec k \vert.
\end{equation}
This solution must satisfy $\vert \vec k \vert^2+k_3^2=0$ between layers, therefore displaying evanescence away from the layers means that $k_3=\pm i k$. To be of finite energy, the order parameter is chosen $\Psi=e^{(i\vec k \cdot \vec x-k|x_3|)}\Psi_{\vec k}$ for a layer at $x_3=0$. Thus between the layers the order parameter does not vanish, which means that the present treatment assumes a metallic medium and not a vacuum. Interestingly $\Psi$ solves $\vec \nabla^2\Psi=0$ above and below but not at the layer itself ($x_3=0$). The Fourier coefficients $\Psi_{\vec k}$ are so far free, and can be determined through the angular momentum properties of the order parameter.

Although the above scheme contains the basic sought ideas, it does not unfold the topological stable solutions that exist in the layered system. To unveil them a dual view of the kinetic energy~\cite{cariglia14} must be obtained, as shown in the next section.
\begin{figure}
\includegraphics[width=\columnwidth]{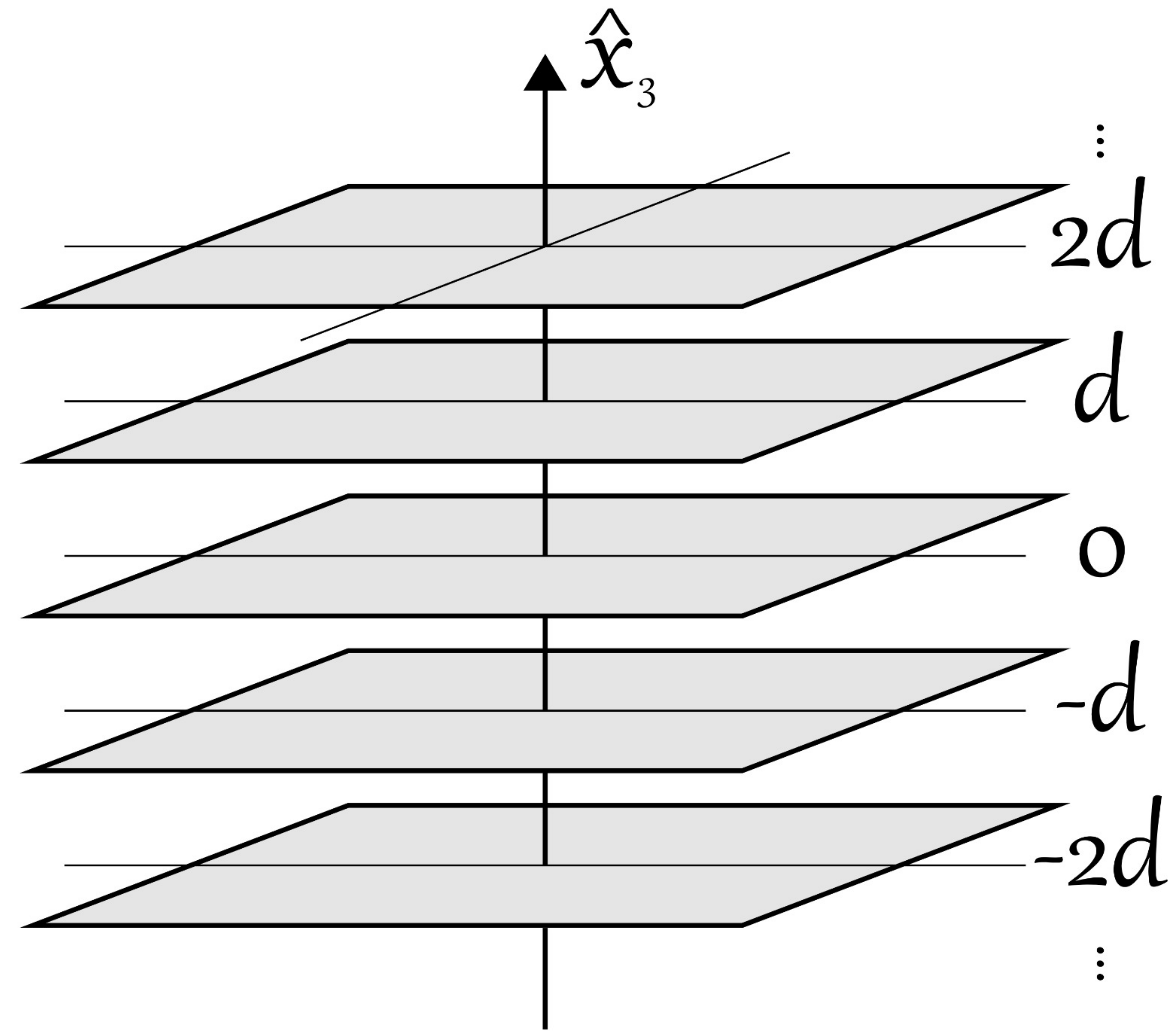}
\caption{Pictorial view of a stack of layers separated by a distance $d$ showing the three dimensional cell associated to the order parameter solution of Eq.(\ref{ml1}).}
\label{f1}
\end{figure}
\begin{figure}
\includegraphics[width=\columnwidth]{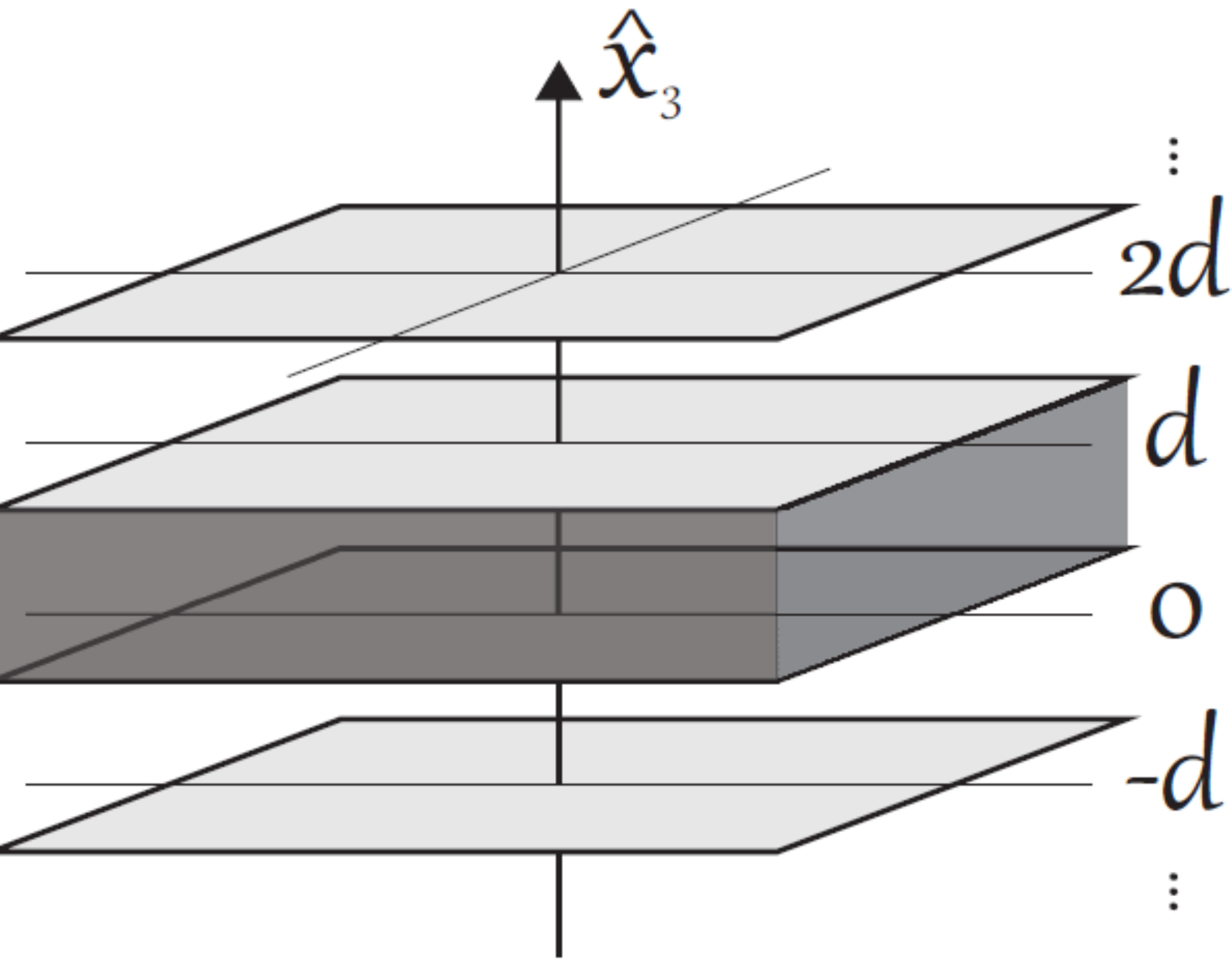}
\caption{Pictorial view of a stack of layers separated by a distance $d$ showing the three dimensional cell associated to the order parameter solution of Eq.(\ref{x30}).}
\label{f2}
\end{figure}

\section{The dual view of the theory}
In this section we construct a dual formulation of the kinetic energy density. To make our demonstration clearer we express three-dimensional indices explicitly such that repeated indices mean a summation according to the Einstein convention.
\begin{equation}
\frac{|\vec{D}\Psi|^2}{2m}=\frac{1}{2m}\left(D_{i}\Psi\right)^{\dag}\delta_{ij}I\left(D_{j}\Psi\right) ,
\end{equation}
and introduce the two-dimensional identity $I=\left(
\begin{array}
[c]{cc}%
1 & 0\\
0 & 1
\end{array}
\right)  $ and  the Clifford algebra associated to the Pauli matrices $\sigma_i$, $i=1,2,3$:
\begin{eqnarray}
\left\{  \sigma_{i},\sigma_{j}\right\}   &  =2\delta_{ij}I,\\
\lbrack\sigma_{i},\sigma_{j}]  &  =2i\varepsilon_{ijk}\sigma_{k}.
\end{eqnarray}
The two-dimensional unity can be expressed in terms of the Pauli matrices,
\begin{equation}
I\delta_{ij}=\sigma_{i}\sigma_{j}-i\varepsilon_{ijk}\sigma_{k}.
\end{equation}
Then the kinetic energy acquires a new form:
\begin{equation}
\frac{|\vec{D}\Psi|^2}{2m}=\frac{1}{2m}\left[|\sigma_{i}%
D_{i}\Psi|  ^{2}-i\varepsilon_{ijk}\left(  D_{i}\Psi\right)  ^{\dag}\sigma_{k}\left(
D_{j}\Psi\right)\right]  \label{ek}.
\end{equation}
Manipulation of the second term of the r.h.s of Eq.(\ref{ek}) gives that,
\begin{eqnarray}
\lefteqn{\varepsilon_{ijk}\left(D_{i}\Psi\right)^{\dag}\sigma_{k}\left(  D_{j}\Psi\right)=}\nonumber \\
&-\frac{\hbar}{2i}\vec{\nabla}.\left[\left(\vec{D}\Psi\right)^{\dag}\times\vec{\sigma}\Psi-\Psi^{\dag}\vec{\sigma}\times\left(\vec{D}\Psi\right)\right]+\nonumber  \\
&+\frac{1}{2}\left[\Psi^{\dag}\vec{\sigma}.\left(\vec{D}\times\vec{D}\Psi\right)-\left(\vec{D}\times\vec{D}\Psi\right)^{\dag}.\vec{\sigma}\Psi\right].\label{sp1}
\end{eqnarray}
The antisymmetry of the Levi-Civita symbol allows us to express
\begin{equation}
\Psi^{\dag}\vec{\sigma}.\left(\vec{D}\times\vec{D}\Psi\right)=-\left(\vec{D}\times \vec{D}\Psi\right)^{\dag}.\vec{\sigma}\Psi=-\frac{\hbar q}{ic}\vec{h}.\Psi^{\dag}\vec{\sigma}\Psi,
\end{equation}
in terms of $\vec{h}$. Finally we reach the dual view of  the kinetic energy:
\begin{eqnarray}
\frac{|\vec{D}\Psi|^2}{2m}&&=\frac{|\vec{\sigma}.\vec{D}\Psi|^{2}}{2m}+\frac{\hbar q}{2mc}\vec{h}.\left(\Psi^{\dag}\vec{\sigma}\Psi\right)+ \nonumber \\
 &&+\frac{\hbar}{4m}\vec{\nabla}.\left[\left(\vec{D}\Psi\right)^{\dag}\times\vec{\sigma} \Psi-\Psi^{\dag}\vec{\sigma}\times\left(\vec{D}\Psi\right)\right]. \label{fk}
\end{eqnarray}
The last term is a total derivative and yet it is a non-vanishing contribution  because of the exponential evanescence of the order parameter away from the layer.
In fact we find that this total derivative term describes the normal state gap associated to the pseudogap.
\subsection{The variational and the first order equations}
The mathematical identity obtained in the previous section is used here to obtain a new perspective of the problem. Using Eq.(\ref{fk}) we can write the free energy of Eq.(\ref{f0}), as follows,
\begin{eqnarray}
F=\int\frac{d^{3}x}{V}\left\{\frac{|\vec{\sigma}.\vec{D}\Psi|^{2}}{2m}+\frac{\hbar q}{2mc}\vec{h}.\left(\Psi^{\dag}\vec{\sigma}\Psi\right)+\right.\nonumber \\
\left.+\frac{\hbar}{4m}\vec{\nabla}.\left[\left(\vec{D}\Psi\right)^{\dag}\times \vec{\sigma}\Psi-\Psi^{\dag}\vec{\sigma}\times\left(\vec{D}\Psi\right)\right]+\frac{\vec{h}^{2}}{8\pi}\right\}. \label{f}
\end{eqnarray}
Taking the variation of Eq.(\ref{f}) for $\delta\Psi^{\dag}$ and $\delta\vec{A}$, we obtain the Ginzburg-Landau equation and Ampere's Law with a new vest:
\begin{eqnarray}
&& \frac{1}{2m}\left( \vec{\sigma}\cdot\vec{D}\right)^2 \Psi=  -\mu_B \vec h \cdot \vec \sigma \Psi \label{gl2} \\
&& \vec \nabla \times \left(\vec h + 4\pi\mu_B\Psi^\dag  \vec{\sigma}\Psi \right) = \nonumber \\
&& =\frac{2\pi q }{mc}\left[\Psi^\dag\vec \sigma
\left (\vec{\sigma}\cdot\vec{D}\Psi\right )+ c.c.\right ]. \label{amp2}
\end{eqnarray}
where $\mu_B=\hbar q/2mc$. The surface term of Eq.(\ref{f}) does not contribute to the above variational equations  because it is a total derivative and $\delta\vec{A}=\delta\Psi^{\dag}=0$ in the boundaries of integration. This second view of the Ampere's law provides the following formulation of the supercurrent:
\begin{equation}
\vec{J}=\frac{q}{2m}\left[\Psi^{\dag}\vec{\sigma}\left(\vec{\sigma}.\vec{D} \Psi\right)+c.c\right]-\frac{\hbar q}{2m}\vec{\nabla}\times\left(\Psi^{\dag}\vec{\sigma}\Psi\right).
\end{equation}
Next we introduce the weak local magnetic field approximation, discussed in the previous section, but now applied to the dual view of the theory. However before doing so, we notice a remarkable feature that can only be seen from the above dual variational equations. Amp\`ere's law, as given by Eq.(\ref{amp2}), is exactly solved by assuming the following first order equations:
\begin{eqnarray}
&&\vec{\sigma}.\vec{D}\Psi=0, \label{fo1}\\
&&\vec{h}=-4\pi\mu_B\Psi^\dag\vec{\sigma}\Psi.
\label{fo2}
\end{eqnarray}
These equations are at the heart of our approach. The solution of Eqs.(\ref{fo1}) and (\ref{fo2}) is also a solution of the variational equations of Eqs.(\ref{gl2}) and (\ref{amp2}) provided that the weak local magnetic field approximation is taken ino account.
In fact Eqs.(\ref{fo1}) and (\ref{fo2}) provide a precise meaning for the weak local magnetic field approximation. A scale transformation $\Psi \rightarrow \varepsilon \Psi$, where $\varepsilon$ is a small parameter, leaves Eq.(\ref{fo1}) invariant. It is Eq.(\ref{fo2}) which sets the scale since under this transformation $\vec h \rightarrow \varepsilon^2 \vec h$. Therefore introducing a solution of the first order equations into Eq.(\ref{gl2}) renders its left hand side null and its right hand side of order $\varepsilon^3$. In conclusion the variational equations are automatically solved by the first order equations in order $\varepsilon^3$. Therefore we solve the first order equations in the lowest possible order instead of  the variational equations. We neglect $\varepsilon^3$ corrections to Eq.(\ref{fo1}) which means to solve $\vec{\sigma}.\vec{\nabla}\Psi=0$ instead. Then Eq.(\ref{fo2}) determines the magnetic field in order $\varepsilon^2$. Following this program the free energy must be obtained under the small magnetic field approximation in order $\varepsilon^3$, to be consistent with the order $\varepsilon^2$ kept in the variational equations. Firstly we show that Eq.(\ref{fo1}) can be written differently if we multiply it by $\vec{\sigma}$:
$\vec{\sigma}\left(\vec{\sigma}.\vec{D}\Psi\right)=-i\vec{\sigma}\times\vec{D}\Psi+\vec{D}\Psi$ The first term in the r.h.s is known as the Rashba interaction~\cite{rashba}. Thus the first order equation (\ref{fo1}) becomes,
\begin{equation}
i\vec{D}\Psi=\vec{\sigma}\times\vec{D}\Psi.
\label{id1}
\end{equation}
Under this relation Eq. (\ref{fk}) can be expressed as,
\begin{equation}
\frac{|\vec{D}\Psi|^2}{2m}=-4\pi\mu_B^2(\Psi^{\dag}\vec{\sigma}\Psi)^2+\frac{\hbar^2}{4m}\vec{\nabla}^2|\Psi|^2\label{fk2}.
\end{equation}
If we discard terms of order $\varepsilon^4$ or superior only the last term has to be kept in the kinetic energy. The field energy is also neglected since it is of order  $\varepsilon^4$ since it is proportional to the square of the magnetic field. Consequently the free energy of Eq.(\ref{f0}) becomes to order $\varepsilon^3$,
\begin{equation}
F=\frac{\hbar^2}{4m}\int\frac{d^3x}{V}\vec{\nabla}^2|\Psi|^2,
\label{kinsup}
\end{equation}
In the next section we obtain  solutions of the first order equations, Eqs.(\ref{fo1}) and (\ref{fo2}), for a single and  multiple layers, respectively. The energy density given by Eq.(\ref{kinsup}) is shown to describe a normal state gap.

\section{The order parameter}
\subsection{Single Layer}
The first order equation for $\Psi$, Eq.(\ref{fo1}), neglecting terms of order $\varepsilon^3$ or higher corresponds to,
\begin{equation}
\vec \sigma \cdot \vec \nabla \Psi=0
\label{signab}
\end{equation}
since $\Psi \propto \varepsilon$, and $\vec{A}\propto \varepsilon^2$. In components it becomes,
\begin{equation}
\left(
\begin{array}
[c]{cc}%
\nabla_3 & \nabla_1-i\nabla_2\\
\nabla_1+i\nabla_2 & -\nabla_3
\end{array}
\right)
\left(\begin{array}[c]{cc}
\psi_u\\
\psi_d
\end{array}\right)
=0.
\label{fo11}
\end{equation}
We seek the solution describing a single layer at $x_3=0$ that vanishes exponentially away from it for $x_3<0$ and $x_3>0$:
\begin{equation}
\Psi=\sum_{\vec{k}\ne 0}e^{-k|x_3|}e^{i\vec{k}.\vec{x}}
\left(\begin{array}[c]{cc}
\psi_u(k)\\
\psi_d(k)
\end{array}\right),
\label{1ll}
\end{equation}
The $\vec k = 0$ component is excluded since it describes a constant homogeneous order parameter,  assumed not to exist here. To obtain a relation between $\psi_u$ and $\psi_d$ components of Eq.(\ref{1ll}) we introduce this solution into Eq.(\ref{fo11}):
\begin{equation}
\psi_d=-i\frac{k_+}{k}\frac{x_3}{|x_3|}\psi_u,
\label{udr}
\end{equation}
where we define,
\begin{equation}
k_\pm \equiv k_1\pm ik_2.
\label{kpm}
\end{equation}
The solution of Eq. (\ref{udr}) only applies outside the layer, similarly to the previous section solution. Finally we write the single layer solution as,
\begin{equation}
\Psi=\sum_{\vec{k}\ne 0}c_{\vec{k}}e^{-k|x_3|}e^{i\vec{k}.\vec{x}}
\left(\begin{array}[c]{cc}
1\\
-i\frac{k_+}{k}\frac{x_3}{|x_3|}
\end{array}\right)
\label{1l}
\end{equation}
Any set of Fourier coefficients  $c_{\vec{k}}$ in Eq. (\ref{1l}) provides a solution of Eq.(\ref{fo1}) valid until order $\varepsilon^3$. We shall determine in this paper these coefficients assuming orbital momentum properties for the order parameter. The single layer solution of Eq.(\ref{1l}) satisfies the following relations:
\begin{eqnarray}\label{slr}
|\Psi|^2=\sum_{\vec{k},\vec{k}'}c_{\vec{k}}^* c_{\vec{k}'}e^{-(k+k')|x_3|}e^{i(\vec{k}'-\vec{k}).\vec{x}}\left(1+\frac{k_-'k_+}{k'k}\right),\nonumber\\
\end{eqnarray}
\begin{eqnarray}
&&\Psi^\dag\sigma_1\Psi= \nonumber \\
&&=i\frac{x_3}{|x_3|}\sum_{\vec{k},\vec{k}'}c_{\vec{k}}^* c_{\vec{k}'}e^{-(k+k')|x_3|}e^{i(\vec{k}'-\vec{k}).\vec{x}}\left(\frac{k_-'}{k'}-\frac{k_+}{k}\right)
\end{eqnarray}
\begin{eqnarray}
&&\Psi^\dag\sigma_2\Psi=\nonumber \\
&&= -\frac{x_3}{|x_3|}\sum_{\vec{k},\vec{k}'}c_{\vec{k}}^* c_{\vec{k}'}e^{-(k+k')|x_3|}e^{i(\vec{k}'-\vec{k}).\vec{x}}\left(\frac{k_-'}{k'}+\frac{k_+}{k}\right),
\end{eqnarray}
\begin{eqnarray}
&&\Psi^\dag\sigma_3\Psi=\nonumber \\
&&=\sum_{\vec{k},\vec{k}'}c_{\vec{k}}^* c_{\vec{k}'}e^{-(k+k')|x_3|}e^{i(\vec{k}'-\vec{k}).\vec{x}}\left(1-\frac{k_-'k_+}{k'k}\right).
\end{eqnarray}
The mean values of the above quantities are calculated through the definition,
\begin{eqnarray}
&&\int{\frac{d^3x}{V}}\left(\cdots\right)\equiv \nonumber \\
&&\int\frac{d^2x}{A}\left(\int_{0^{+}}^{\infty}\frac{dx_3}{l}\left( \cdots \right)+\int_{-\infty}^{0^{-}}\frac{dx_3}{l}\left( \cdots \right)\right),
\label{avdef}
\end{eqnarray}
where $l$ is an arbitrary length by taking $V=Al$. Consider that,
\begin{equation}
\int \frac{d^2x}{A} e^{i(\vec{k}-\vec{k}')} = \delta_{\vec{k}\vec{k}'}, \label{delta}
\end{equation}
where $A$ is the area of the unitary cell, one obtains that,
\begin{eqnarray}
\int{\frac{d^3x}{V}}|\Psi|^2=\frac{2}{l}\sum_{\vec{k}\ne 0}\frac{|c_{\vec{k}}|^2}{k},\label{mp2}
\end{eqnarray}
\begin{eqnarray}
\int{\frac{d^3x}{V}}\Psi^\dag\sigma_1\Psi = 0\label{ms1},
\end{eqnarray}
\begin{eqnarray}
\int{\frac{d^3x}{V}}\Psi^\dag\sigma_2\Psi = 0\label{ms2},
\end{eqnarray}
\begin{eqnarray}
\int{\frac{d^3x}{V}}\Psi^\dag\sigma_3\Psi = 0\label{ms3},
\end{eqnarray}
\begin{eqnarray}
\int{\frac{d^3x}{V}}\nabla^2|\Psi|^2 = \frac{8}{l}\sum_{\vec{k}\ne 0}k|c_{\vec{k}}|^2,\label{ms4}
\end{eqnarray}
The mean values vanish for different reasons, while
Eqs.(\ref{ms1}) and (\ref{ms2}) are null because contributions above and below the layer are equal, and the factor $\frac{x_3}{|x_3|}$ flips their sign, Eq.(\ref{ms3}) is zero because of integration along the layer that renders $\vec{k}=\vec{k}'$.

\subsection{Multiple layers}
The general solution for a stack of layers separated by a distance $d$ is given by,
\begin{equation}
\Psi=\sum_{\vec{k}\ne 0,n}{c_{\vec{k}}^{(n)}e^{-k|x_3-nd|}e^{i\vec{k}.\vec{x}}\left( \begin{array}[c]{cc}
{1}\\
-i\frac{k_+}{k}\frac{x_3-nd}{|x_3-nd|}
\end{array} \right)}.
\label{ml}
\end{equation}
Because all the layers are assumed to be identical, $c_{\vec{k}}^{(n)}=c_{\vec{k}}$. We study two different ways to sum over the layers that give distinct but equivalent results, valid for the regions $0<x_3<d$ and $-d<x_3<d$, as shown in Figs.(\ref{f1}) and (\ref{f2}), respectively. Basically they differ on their treatment of the sign provided by $\frac{x_3}{|x_3|}$, above and below the corresponding region. The solution that applies to $0<x_3<d$ takes into account that,
\begin{eqnarray}
&& \sum_{n=-\infty}^{+\infty}e^{-k|x_3-nd|}=\frac{\cosh\left[k\left(x_3-\frac{d}{2}\right)\right]}{\sinh\left(\frac{kd}{2}\right)}\label{sum1},\\
&& \sum_{n=-\infty}^{+\infty}\frac{x_3-nd}{|x_3-nd|}e^{-k|x_3-nd|}=\frac{\sinh\left[k\left(x_3-\frac{d}{2}\right)\right]}{\sinh\left(\frac{kd}{2}\right)}\label{sum2},
\end{eqnarray}
which yield the following multilayer solution:
\begin{equation}
\Psi=\sum_{\vec{k}\ne 0}c_{\vec{k}}{\frac{e^{i\vec{k}.\vec{x}}}{\sinh(\frac{kd}{2})}\left( \begin{array}[c]{cc}
\cosh\left[k\left(x_3-\frac{d}{2}\right)\right]\\
i\frac{k_+}{k}\sinh\left[k\left(x_3-\frac{d}{2}\right)\right]
\end{array} \right)}
\label{ml1}
\end{equation}
The validity of Eq.(\ref{ml1}) is restricted to the region $0<x_3<d$. For the region shown in Fig.(\ref{f2}), namely $-d<x_3<d$, the sum over the layers consider the  $x_3=0$ layer separately, $ \sum_{n\neq 0}e^{-k|x_3-nd|}+e^{-k|x_3|}$ and $\sum_{n\neq 0}\frac{x_3-nd}{|x_3-nd|}e^{-k|x_3-nd|}+e^{-k|x_3|}\frac{x_3}{|x_3|}$, to obtain that,
\begin{widetext}
\begin{equation}
\Psi=\sum_{\vec{k}\ne 0}c_{\vec{k}}e^{i\vec{k}.\vec{x}}\left[ e^{-k|x_3|}\left( \begin{array}[c]{cc}
1\\
-i\frac{k_+}{k}\frac{x_3}{|x_3|}
\end{array} \right)+\frac{2}{(e^{kd}-1)}\left( \begin{array}[c]{cc}
\cosh(kx_3)\\
i\frac{k_+}{k}\sinh(kx_3)
\end{array} \right)\right],
\label{x30}
\end{equation}
\end{widetext}
In this last expression the single layer solution of Eq.(\ref{1l}) is easily retrieved by taking the limit $d\rightarrow \infty$.
To obtain the mean values, similarly  to the the single layer solution, requires that the mean value be taken over a well defined unit cell since along the direction perpendicular to the layers $l=d$: $\int{\frac{d^3x}{V}} (\cdots)\equiv \int_0^d{\frac{dx_3}{d}}\int{\frac{d^2x}{A}}(\cdots)$. Then one obtains that,
\begin{eqnarray}
\int{\frac{d^3x}{V}}|\Psi|^2 =\frac{2}{d}\sum_{\vec{k}\ne 0}\frac{|c_{\vec{k}}|^2}{k}\coth\left(\frac{kd}{2}\right),\label{mp2m}
\end{eqnarray}
\begin{eqnarray}
\int{\frac{d^3x}{V}}\Psi^\dag\sigma_1\Psi = 0\label{mms1},
\end{eqnarray}
\begin{eqnarray}
\int{\frac{d^3x}{V}}\Psi^\dag\sigma_2\Psi = 0\label{mms2},
\end{eqnarray}
\begin{eqnarray}
\int{\frac{d^3x}{V}}\Psi^\dag\sigma_3\Psi = \sum_{\vec{k}\ne 0}\frac{|c_{\vec{k}}|^2}{\sinh^2(\frac{kd}{2})}\label{mms3},
\end{eqnarray}
\begin{eqnarray}
\int{\frac{d^3x}{V}}\nabla^2|\Psi|^2 =  \frac{8}{d}\sum_{\vec{k}\ne 0} k |c_{\vec{k}}|^2 \coth(\frac{kd}{2}) \label{mms4}.
\end{eqnarray}
In the limit $d\rightarrow \infty$, Eqs.(\ref{mp2m}), (\ref{mms3}) and (\ref{mms4}) yield their single layer counterparts, given by Eqs.(\ref{mp2}), (\ref{ms3})  and (\ref{ms4}), respectively. Notice that Eq.(\ref{mms3}) is non zero in contrast with the single layer case. Also Eqs.(\ref{mms1}) and (\ref{mms4}) have a $\frac{1}{d}$ factor due to the volume integration, which is also present in Eqs.(\ref{ms1}) and (\ref{ms4}), as the $1/l$ factor.
\begin{figure}
\includegraphics[width=\columnwidth]{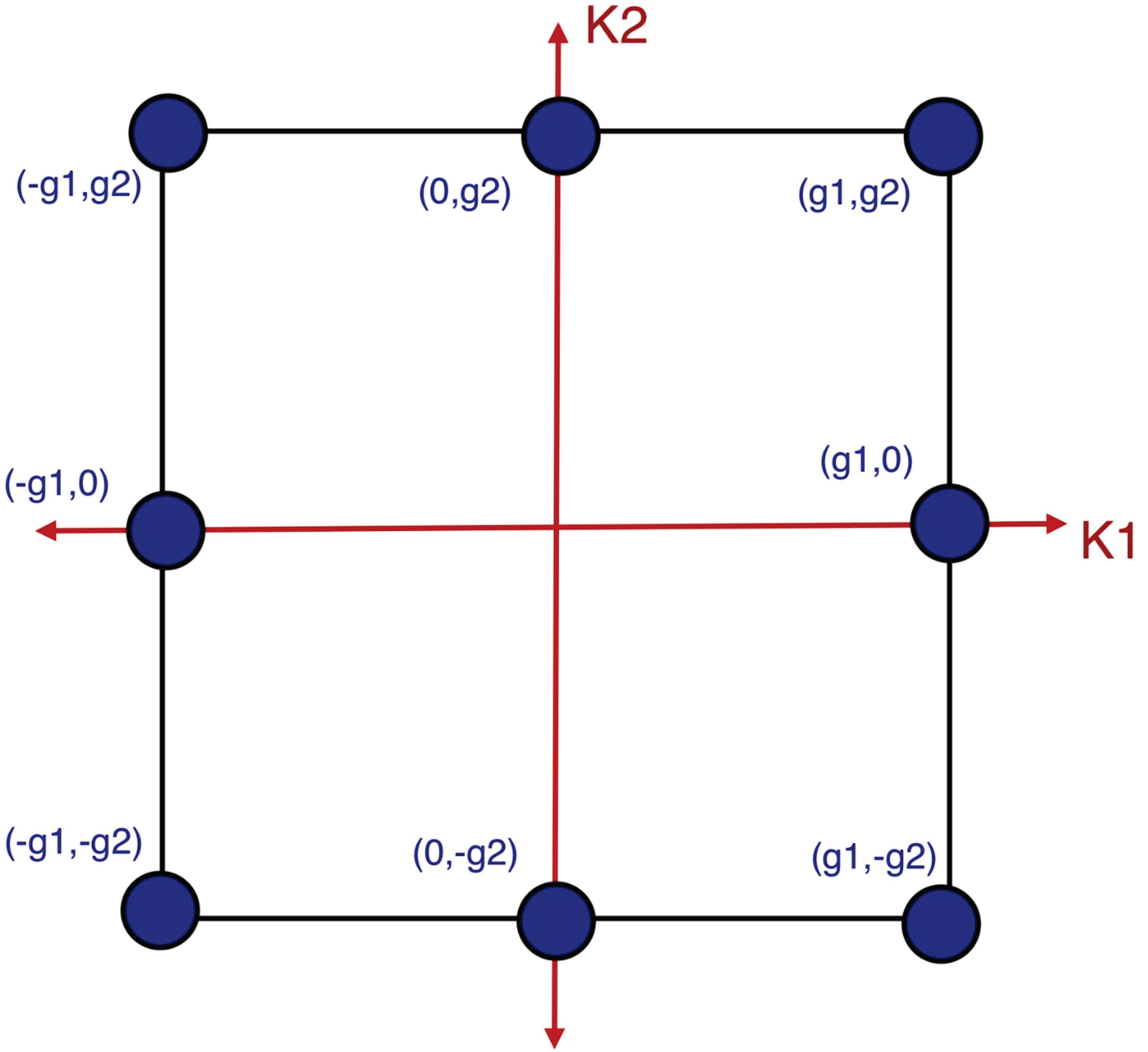}
\caption{The restricted Fourier space contains eight points in a rectangular unitary cell of sides $L_1$ and $L_2$, where $g_1=\frac{2\pi}{L_1}$ and $g_2=\frac{2\pi}{L_2}$.}
\label{fexpan}
\end{figure}

\section{The broken symmetries}
A remarkable difference between the single and the multi layer solutions is that in the latter case the mean value of the local magnetic field is non zero and points perpendicularly to the layers, regardless of the choice of the coefficients $c_{\vec{k}}$, according to Eqs.(\ref{fo2}), (\ref{mms1}),(\ref{mms2}), and (\ref{mms3}):
\begin{eqnarray}
\int{\frac{d^3x}{V}} \vec h\left(\Psi\right) = -4\pi\mu_B \sum_{\vec{k}\ne 0}\frac{|c_{\vec{k}}|^2}{\sinh^2(\frac{kd}{2})}\hat x_3\label{mean3}.
\end{eqnarray}
This is a clear consequence of the breaking of the space reversal symmetry  in the multi layer case. Nevertheless the multi layer solution is derived from the first order equations, defined by  Eqs.(\ref{fo1}) and (\ref{fo2}), which do not break this symmetry. Therefore another solution is expected such  that  the two possible ways of breaking the space reversal symmetry must be possible. Indeed this other solution $\Psi'$ is obtained by expressing Eq.(\ref{udr}) as,
\begin{equation}
\psi'_u=i\frac{k_-}{k}\frac{x_3}{|x_3|}\psi'_d.
\label{nuds}
\end{equation}
The corresponding single layer solution is,
\begin{equation}
\Psi'=\sum_{\vec{k}\ne 0}\tilde{c}_{\vec{k}}e^{-k|x_3|}e^{i\vec{k}.\vec{x}}
\left(\begin{array}[c]{cc}
i\frac{k_-}{k}\frac{x_3}{|x_3|}\\1
\end{array}\right)
\label{t1l}
\end{equation}
and the multi layer is,
\begin{equation}
\Psi'=\sum_{\vec{k}\ne 0}\tilde{c}_{\vec{k}}{\frac{e^{i\vec{k}.\vec{x}}}{\sinh(\frac{kd}{2})}\left( \begin{array}[c]{cc}
-i\frac{k_-}{k}\sinh\left[k\left(x_3-\frac{d}{2}\right)\right]\\
\cosh\left[k\left(x_3-\frac{d}{2}\right)\right]
\end{array} \right)}
\label{tml1}
\end{equation}
The mean values are given by,
\begin{eqnarray}
\int{\frac{d^3x}{V}}|\Psi'|^2 =\frac{2}{d}\sum_{\vec{k}\ne 0}\frac{|\tilde c_{\vec{k}}|^2}{k}\coth\left(\frac{kd}{2}\right),\label{tp2m}
\end{eqnarray}
\begin{eqnarray}
\int{\frac{d^3x}{V}}\Psi'^\dag\sigma_1\Psi' = 0\label{tms1},
\end{eqnarray}
\begin{eqnarray}
\int{\frac{d^3x}{V}}\Psi'^\dag\sigma_2\Psi' = 0\label{tms2},
\end{eqnarray}
\begin{eqnarray}
\int{\frac{d^3x}{V}}\Psi'^\dag\sigma_3\Psi' = -\sum_{\vec{k}\ne 0}\frac{|\tilde c_{\vec{k}}|^2}{\sinh^2(\frac{kd}{2})}\label{tms3},
\end{eqnarray}
\begin{eqnarray}
\int{\frac{d^3x}{V}}\nabla^2|\Psi'|^2 =  \frac{8}{d}\sum_{\vec{k}\ne 0} k |\tilde c_{\vec{k}}|^2 \coth(\frac{kd}{2}) \label{tms4}.
\end{eqnarray}
The mean local magnetic field is found to point oppositely to the solution used in Eq.(\ref{mean3}):
\begin{eqnarray}
\int{\frac{d^3x}{V}} \vec h\left(\Psi' \right) = 4\pi\mu_B \sum_{\vec{k}\ne 0}\frac{|\tilde c_{\vec{k}}|^2}{\sinh^2(\frac{kd}{2})}\hat x_3\label{tmean3}.
\end{eqnarray}
The existence of solutions with $\int\frac{d^3x}{V}\vec h$ in opposite directions, given by Eqs.(\ref{ml1}) and (\ref{tml1}), in case of multi layers is suggestive of the coexistence of these solutions separated by domain walls. Their free energy has the same form,
\begin{eqnarray}
F= 4\frac{\hbar^2}{md^2}\sum_{\vec{k}\ne 0}|c_{\vec{k}}|^2 \frac{kd}{2} \coth(\frac{kd}{2}),
\end{eqnarray}
as obtained from Eqs.(\ref{kinsup}), (\ref{mms4}) and (\ref{tms4}), respectively. The reverted magnetic field solution has  $|c_{\vec{k}}|^2$ replaced by $|\tilde c_{\vec{k}}|^2$ in the above expression. In case that $|\tilde c_{\vec{k}}|=|c_{\vec{k}}|$ these two solutions are degenerate in energy and favor the onset of domain walls. We will not treat such coexistence here and concentrate our study only in the $\Psi$ solution.

A remarkable feature of the present approach is the existence of two kinds of supercurrent densities, namely, volumetric and superficial, located between layers and within the layers, respectively. The single layer results for $\Psi^{\dagger}\sigma_1\Psi$ and  $\Psi^{\dagger}\sigma_1\Psi$, given by Eqs.(\ref{slr}), show that above and below the $x_3=0$ layer such values flip sign because of the factor $\frac{x_3} {|x_3|}$. Consequently Eq.(\ref{fo2}) show that the magnetic fields $h_1$ and $h_2$, parallel to this layer, are discontinuous across the $x_3=0$ layer. It is a straightforward consequence of Maxwell's boundary condition that there is a superficial supercurrent configuration given by,
\begin{equation}
\widehat{x}_3\times\left[\vec{h}(0^+)-\vec{h}(0^-)\right]=\frac{4\pi}{c}\vec{J}_{sup}.
\label{cconfig}
\end{equation}
These boundary conditions under a fixed $\vec{J}_{sup}$ reveal the  mechanism of time reversal and parity breaking. The time reversal symmetry transformation is given by $\vec{h}\rightarrow -\vec{h}$ in Eq.(\ref{cconfig}), which is not invariant under this transformation. The parity symmetry transformation, given by $\widehat{x}_3\rightarrow -\widehat{x}_3$ in Eq.(\ref{cconfig}), is also not invariant. Only the product of the time reversal and parity transformations (PT) remains invariant under the presence of the surface supercurrent. We find that the surface supercurrent leads to the onset of a charge density wave whose properties will be studied elsewhere.

\begin{figure}
\includegraphics[width=\columnwidth]{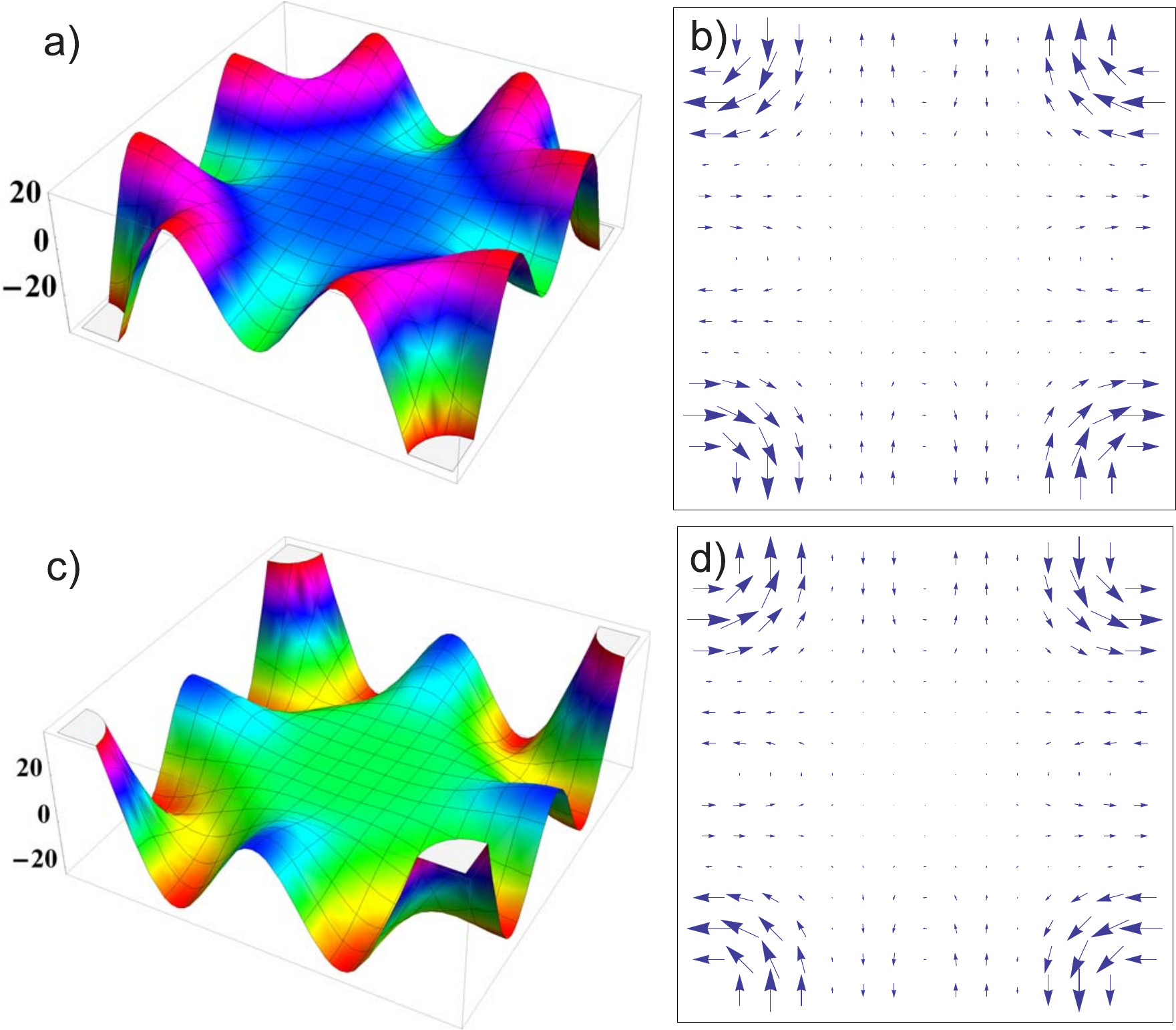}
\caption{ The subplots (a),(c) show  the local magnetic field, $h_3$,, and the subplots (b), (d) the surface supercurrent, $J_s$, for the $J_3=\frac{1}{2}\hbar$ and $J_3=-\frac{1}{2}\hbar$ states, respectively.}
\label{j3half}
\end{figure}
\begin{figure}
\includegraphics[width=\columnwidth]{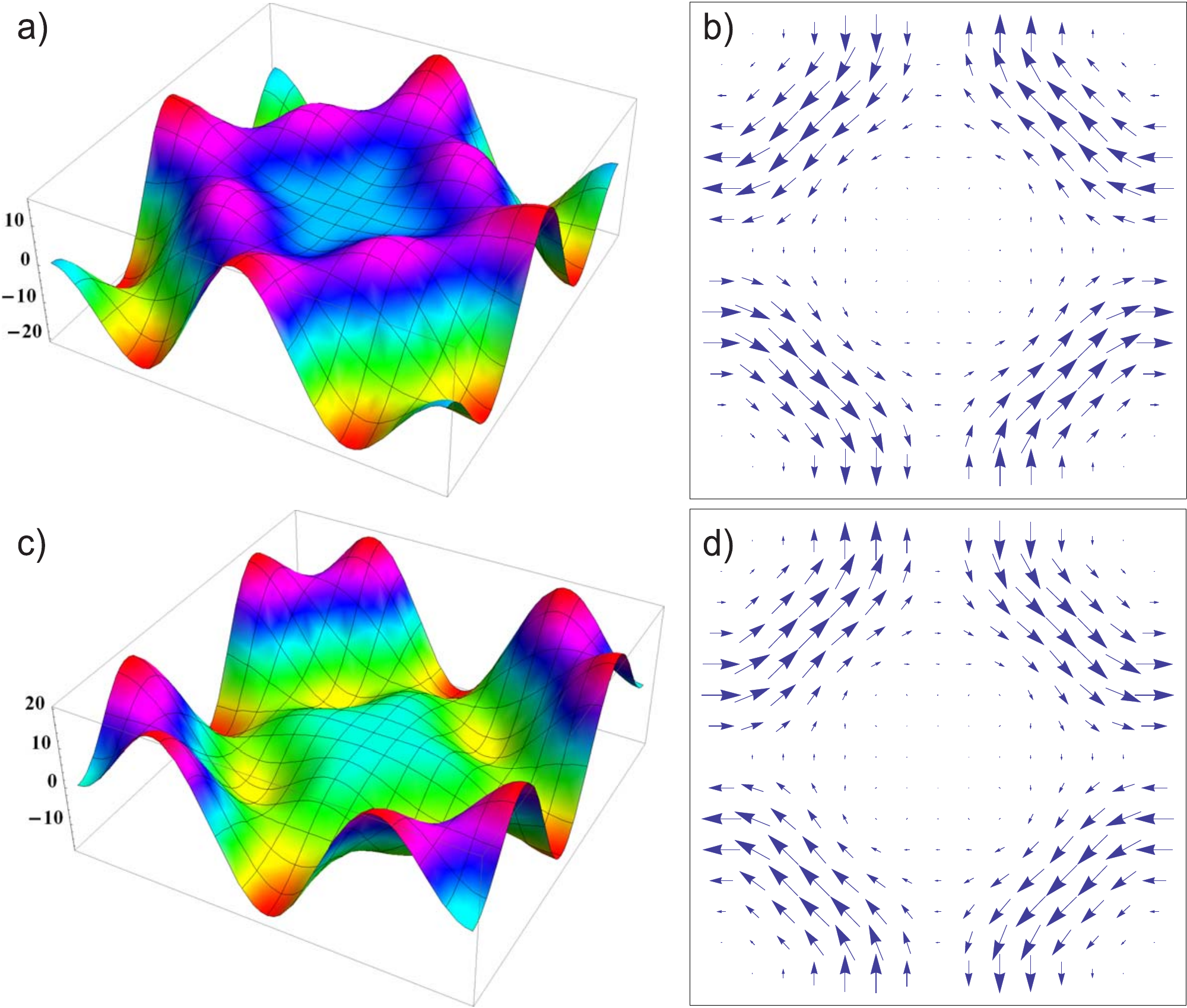}
\caption{The subplots (a),(c) show  the local magnetic field, $h_3$,, and the subplots (b), (d) the surface supercurrent, $J_s$, for the $J_3=\frac{3}{2}\hbar$ and $J_3=-\frac{3}{2}\hbar$ states, respectively.}
\label{j3threehalf}
\end{figure}
\begin{figure}
\includegraphics[width=\columnwidth]{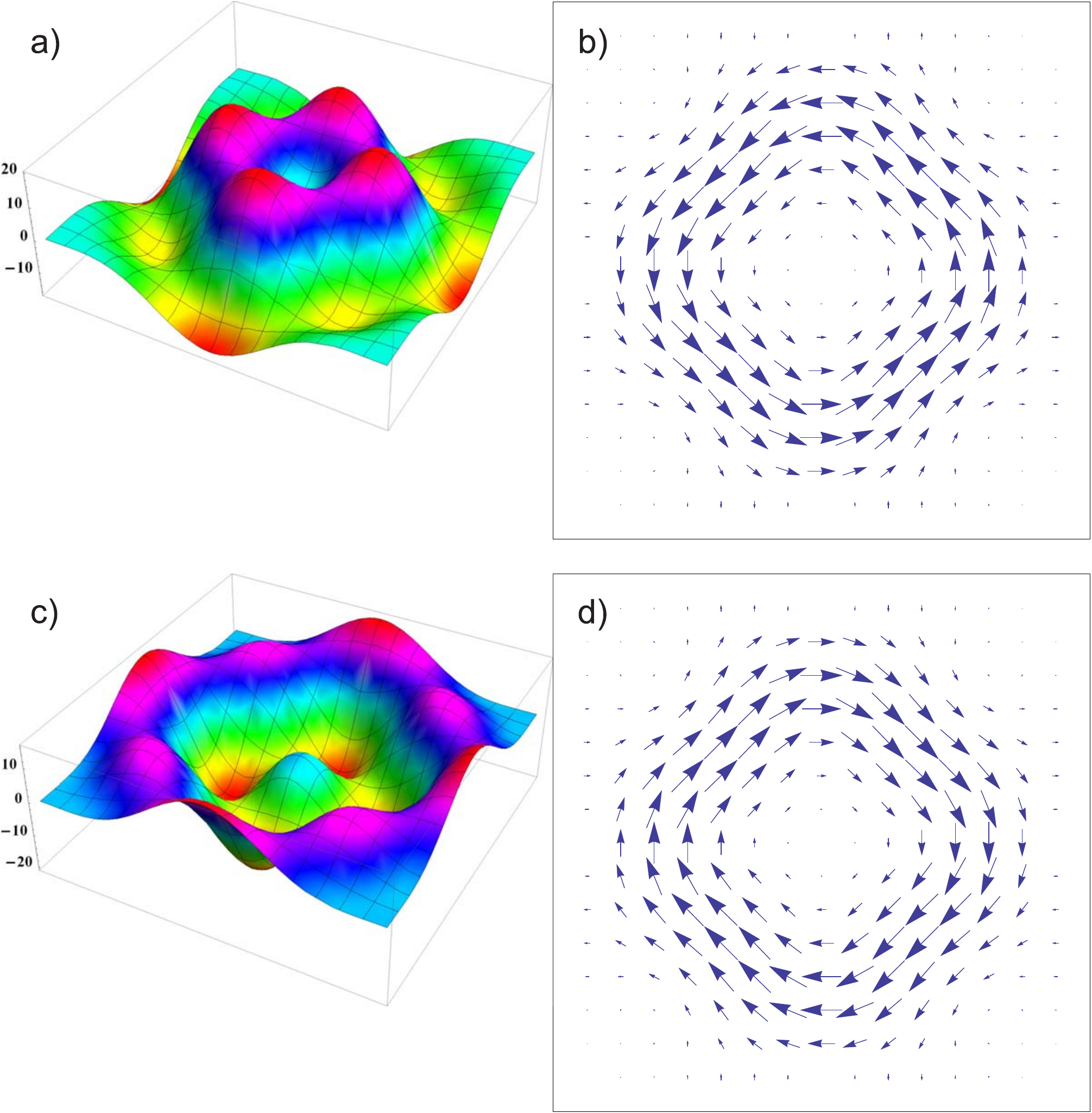}
\caption{The subplots (a),(c) show  the local magnetic field, $h_3$,, and the subplots (b), (d) the surface supercurrent, $J_s$, for the $J_3=\frac{5}{2}\hbar$ and $J_3=-\frac{5}{2}\hbar$ states, respectively.}
\label{j3fivehalf}
\end{figure}

\section{The angular momentum states}
The layered structure  has a preferable direction, defined by the $x_3$ axis, an so, we take from the total angular momentum,
\begin{eqnarray}
\vec{J}=\vec{x}\times\frac{\hbar}{i}\nabla+\frac{\hbar}{2}\vec{\sigma},
\label{dfo}
\end{eqnarray}
the component $J_3$ to determine the coefficients $c_{\vec{k}}$. It happens that up to order $\varepsilon^3$ Eq.(\ref{fo1}) becomes $\vec{\sigma}\cdot\vec{\nabla}\Psi=0$ and it holds that
$[J_i,\vec \sigma \cdot \vec \nabla]=0$, $i=1,2,3$. Therefore we study  $J_3\Psi$, which is also a solution since $\vec \sigma \cdot \vec \nabla(J_3\Psi)=0$.
Although we are not treating a quantum mechanical problem here, the present Hilbert space considerations are useful since the order parameter can be regarded as a superposition of angular momentum states $m$,
\begin{eqnarray}\label{j3e}
\Psi=\sum_{m}{\alpha}_m\Psi_m, \mbox{where}\; J_3\Psi_m=\hbar\left(m+\frac{1}{2}\right)\Psi_m,
\end{eqnarray}
where $\alpha_m$ are arbitrary coefficients. For simplicity we shall not study here admixtures and only pure $m$ states, which means that all but one of the coefficients $\alpha_m$ vanish. Consider the orbital angular momentum component $l_3=x_1p_2-x_2p_1$, which admits a position representation, $p_i=\frac{\hbar}{i}\frac{\partial}{\partial x_i}$, and also a wave number representation, $x_i=i\frac{\partial}{\partial k_i}$ and $p_i=\hbar k_i$.
For instance, apply the position space representation, $l_3(x)$,  to the up component $\psi_u$ of Eq.(\ref{1l}): $l_3(x)\psi_u(x) =
\hbar\sum_{\vec{k}}c_{\vec{k}}\left(x_1 k_2-x_2 k_1\right)e^{i\vec{k}\cdot\vec{x}-k|x_3|}$.
In wave number space representation it holds that,
$l_3\psi_u = \sum_{\vec{k}}c_{\vec{k}}\left[-l_3(k)\,e^{i\vec{k}\cdot\vec{x}-k|x_3|}\right]=\sum_{\vec{k}}\left[ l_3(k)\,c_{\vec{k}}\right ]e^{i\vec{k}\cdot\vec{x}-k|x_3|}$, the last expression follows since a sum over a total derivative is null. Thus $l_3$ can be thought of as directly acting on the coefficients $c_{\vec{k}}$.
In wave number space the eigenvector eigenvalue problem for the $l_3$ operator is easily soluble: $l_3k_\pm^m=\pm m\hbar k_\pm^m$, thus $(k_\pm)^m$ and $\pm \hbar m$ are the sought eigenstates and eigenvalues, respectively. For any two functions, $f$ and $g$, in wave number space, it holds that $l_3\left[f(k)g(k)\right]=[l_3f(k)]g(k)+f(k)[l_3g(k)]$.
For instance, $l_3(k^2)=l_3(k_{+}k_{-})=0$. There are two sets of eigenvectors of $J_3$: $c_{\vec{k}}=\varepsilon\left (\frac{k_{\pm}}{k}\right)^m$. The normalization $1/k^m$ is introduced to avoid growth of the coefficients $c_{\vec{k}}$ for large $k$, and  the multiplicative constant $\varepsilon$ is arbitrary. The total angular momentum eigenvector problem becomes, $J_3\Psi_{m,\pm}=\hbar(\pm m+\frac{1}{2})\Psi_{m,\pm}$, but the property $\left(\frac{k_-}{k}\right)^m=\left(\frac{k_+}{k}\right)^{-m}$
shows that there is truly just one set of eigenvectors,
$ \Psi_m \equiv \Psi_{m,+}, \mbox{since} \; \Psi_{m,-}=\Psi_{-m,+}$,
as long as $m$ ranges from positive to negative values. Writing $k_{\pm}=ke^{\pm i \phi}$ gives that $c_{\vec{k}}=\varepsilon e^{\pm i m\phi}$ and shows that the two signs, $\pm m$, are associated to rotations in opposite directions by angles $\pm m\phi$.
Therefore  the order parameters becomes,
\begin{equation}
\Psi_{m}=\varepsilon \sum_{\vec{k}\ne 0}{\left( \frac{k_{+}}{k} \right)^m\frac{e^{i\vec{k}.\vec{x}}}{\sinh(\frac{kd}{2})}\left( \begin{array}[c]{cc}
\cosh\left( k\overline{x}_3 \right)\\
i\frac{k_+}{k}\sinh\left( k\overline{x}_3 \right)
\end{array} \right)},
\label{pmpsi}\\
\end{equation}
where $\overline{x}_3=x_3-d/2$.
Interestingly  the mean values for the angular momentum order parameter, $\Psi_m$, are independent of $m$, since $|c_{\vec{k}}|^2=\varepsilon^2$.
\begin{eqnarray}
&&\int{\frac{d^3x}{V}}|\Psi_m|^2 =\varepsilon^2\frac{2}{d}\sum_{\vec{k}\ne 0}\frac{1}{k}\coth\left(\frac{kd}{2}\right),\label{cp2m}
\end{eqnarray}
\begin{eqnarray}
&&\int{\frac{d^3x}{V}}\Psi_m^\dag\sigma_1\Psi_m = 0\label{cms1},
\end{eqnarray}
\begin{eqnarray}
&&\int{\frac{d^3x}{V}}\Psi_m^\dag\sigma_2\Psi_m = 0\label{cms2},
\end{eqnarray}
\begin{eqnarray}
&&\int{\frac{d^3x}{V}}\Psi_m^\dag\sigma_3\Psi_m = \varepsilon^2 \sum_{\vec{k}\ne 0}\frac{1}{\sinh^2(\frac{kd}{2})}\label{cms3},
\end{eqnarray}
\begin{eqnarray}
&&\int{\frac{d^3x}{V}}\nabla^2|\Psi_m|^2 =  \varepsilon^2\frac{8}{d}\sum_{\vec{k}\ne 0} k \coth(\frac{kd}{2}) \label{cms4}.
\end{eqnarray}
We find quite remarkable that the mean magnetic field, which relies on Eq.(\ref{fo2}), and the above Eqs.(\ref{cms1}), (\ref{cms2}), and (\ref{cms3}), is angular momentum ($m$) independent. The same holds for the free energy which defines the normal state gap, given by Eqs.(\ref{kinsup}) and (\ref{cms4}), also shown to be $m$ independent. The single layer limit with definite angular momentum $m$ is simply retrieved from the above equations at the limit $d \rightarrow \infty$ once noticed that the behavior $1/d$ is just a consequence of the volume $V=A \,d$, and so, in case of the single layer, $d$ can be replaced by an arbitrary length $l$. We shall not treat here admixtures of different $m$ states and concentrate only in the pure cases.

\section{The tetragonal lattice}
To reach further understanding of the order parameter we simplify matters by restricting the Fourier components, defined as $k_1\equiv g_1n_1$, $k_2 \equiv g_2n_2$ ($g_1=\frac{2\pi}{L_1}, g_2=\frac{2\pi}{L_2}$) for a orthorombic cell $A=L_1 L_2$, to a small set $n_1=-1,0,1$ and $n_2=-1,0,1$ with $n_1=n_2=0$ excluded (no homogeneous state). Then the sum over $\vec{k}$ is restricted to the eight points of Fig.(\ref{fexpan}).
We shall also restrict our analysis to the tetragonal symmetry which means that $L_1=L_2=L$ and $g_1=g_2=g=\frac{2\pi}{L}$ and in this case unit cell $A=L^2$ will be interpreted as that of the checkerboard pattern. Then the order parameter becomes,

\begin{widetext}
\begin{eqnarray}
&&\Psi_{m}=\varepsilon{\frac{2}{\sinh(\frac{gd}{2})}\left( \begin{array}[c]{cc}
\left[e^{i\frac{m\pi}{2}}\cos\left(\frac{m\pi}{2}-gx_1\right)+\cos\left(\frac{m\pi}{2}+gx_2\right)\right]\cosh\left( g\overline{x}_3 \right)\\
\left[e^{i\frac{m\pi}{2}}\sin\left(\frac{m\pi}{2}-gx_1\right)-i\sin\left(\frac{m\pi}{2}+gx_2\right)\right]\sinh\left( g\overline{x}_3 \right)
\end{array} \right)}\nonumber\\
&&+\varepsilon\frac{2}{\sinh(\frac{gd}{\sqrt{2}})}\left( \begin{array}[c]{cc}
\{e^{i\frac{m\pi}{4}}\cos\left[\frac{m\pi}{2}+g(-x_1+x_2)\right]+e^{-i\frac{m\pi}{4}}\cos\left[\frac{m\pi}{2}+g(x_1+x_2)\right]\}\cosh\left(\sqrt{2}g\overline{x}_3 \right)\\
-\{e^{i\frac{(m-1)\pi}{4}}\sin\left[\frac{m\pi}{2}+g(-x_1+x_2)\right]+e^{-i\frac{(m-1)\pi}{4}}\sin\left[\frac{m\pi}{2}+g(x_1+x_2)\right]\}\sinh\left(\sqrt{2}g\overline{x}_3 \right)
\end{array} \right),
\label{psipm}
\end{eqnarray}
\end{widetext}
Eq.(\ref{psipm}) describes the multilayer order parameter with restricted Fourier components in a well defined angular momentum state, as defined by $J_3$ whose eigenvalues are given by Eq.(\ref{j3e}). Because of the restricted Fourier space it has the property that $\Psi_{m+8}=\Psi_m$. Again we notice that mean values acquire very simple forms, which are $m$ independent:
\begin{eqnarray}
\int{\frac{d^3x}{V}}|\Psi_{m}|^2=
\varepsilon^2\frac{4}{gd}\left[2\coth\left(\frac{gd}{2}\right)+\sqrt{2}\coth\left(\frac{\sqrt{2}gd}{2}\right)\right],\nonumber\\
\end{eqnarray}
\begin{eqnarray}
\int{\frac{d^3x}{V}}{\Psi_{m}}^\dag\sigma_1\Psi_{m} = 0,
\end{eqnarray}
\begin{eqnarray}
\int{\frac{d^3x}{V}}{\Psi_{m}}^\dag\sigma_2\Psi_{m} = 0,
\end{eqnarray}
\begin{eqnarray}
\int{\frac{d^3x}{V}}{\Psi_{m}}^\dag\sigma_3\Psi_{m} =
\varepsilon^2 4\left[\frac{1}{\sinh^2\left(\frac{gd}{2}\right)}+\frac{1}{\sinh^2\left(\frac{\sqrt{2}gd}{2}\right)}\right],\nonumber \\
\end{eqnarray}
\begin{eqnarray}
&&\int{\frac{d^3x}{V}}\nabla^2|\Psi_{m}|^2 = \nonumber\\
&&\varepsilon^2 \frac{32g}{d}\left[2\coth\left(\frac{gd}{2}\right)+\sqrt{2}\coth\left(\frac{\sqrt{2}gd}{2}\right)\right].
\end{eqnarray}
They are straightforwardly obtained from Eqs.(\ref{cp2m}), (\ref{cms1}), (\ref{cms2}), (\ref{cms3}), and (\ref{cms4}), once noticed that there are four identical contributions coming from $\vec k$ along the axis and other four coming from the diagonals of Fig.(\ref{fexpan}), and each set contributes with $k=g$ and $k=\sqrt{2}g$, respectively.

\section{The experimental values}
We show here that the normal (above the homogeneous) state gap can be determined by assumption of the experimental value for the local magnetic field between layers.
To completely determine the multi layer order parameter, given by Eq.(\ref{psipm}), the constant $\varepsilon$ must be obtained. This  indefiniteness is also present in the free energy obtained from Eqs.(\ref{kinsup}) and (\ref{cms4}).
\begin{equation}
F=\frac{\hbar^2}{4m}\varepsilon^2\frac{32g}{d}\left[2coth\left(\frac{gd}{2}\right)+\sqrt{2}coth\left(\frac{gd}{\sqrt{2}}\right) \right].
\label{gdensf}
\end{equation}
The so called infrared limit, or limit of very small wave number, of the above expression shows undoubtedly the existence of a normal state gap density:
\begin{equation}
F(L \rightarrow \infty) \rightarrow 48 \varepsilon^2 \frac{\left(\hbar/d\right )^2}{m}.
\label{gap}
\end{equation}
As previously discussed, the constant $\varepsilon$ is determined by small magnetic field condition set by experimental limit of NMR/NQR~\cite{strassle08,strassle11} and  $\mu$SR~\cite{macdougall08,sonier09} experiments. To establish this connection between theory and experiment we obtain the mean value of the local magnetic field from Eq.(\ref{fo2}). Interestingly for a $\Psi_m$ state $\int \frac{d^3x}{V}\vec{h} =0$ for a single layer while for a stack of layers,
 \begin{equation}
\int \frac{d^3x}{V}\vec{h}=-16\pi\mu_{\beta}\varepsilon^2\left[\frac{1}{\sinh^2\left(\frac{\pi d}{L}\right)}+\frac{1}{\sinh^2\left(\frac{\sqrt{2}\pi d}{L}\right)}\right]\hat x_3.
\label{ulimit}
\end{equation}
The vanishing of the mean field for a single layer but not for multi layers is a consequence of a "solenoid" effect. The mean field of the single supercurrent loop vanishes but for a stack of current loops it does not vanish, since it is non-zero inside and zero outside. Because Eq.(\ref{ulimit}) applies for any $m$ this solenoid interpretation is not directly related to circulating supercurrents within the layers.
Next we take that this experimental threshold sets the theoretical  mean value:
\begin{equation}
\vert \, \int \frac{d^3x}{V}\vec{h} \; \vert =h_{exp},
\label{ulimit2}
\end{equation}
where we choose to work with $h_{exp}=0.01\; \mbox{Gauss}$. The ratio $d/L$ must also be known to determine the value of  $\varepsilon$ , where $d$ is  the distance between two consecutive layers and $L$ defines the size of the tetragonal unit cell within a layer. For numerical purposes we have in mind the compound $YBa_2Cu_3O_{7-0.08}$ as this material presents the checkerboard pattern~\cite{zahirul04} with $L=4a=1.6 \; nm$, as we take $a = 0.4 \; \mbox{nm}$ and $d = 1.2 \; \mbox{nm}$. Thus this ratio is given by,
\begin{equation}
\frac{d}{L}= 0.75.
\label{ratio}
\end{equation}
In convenient units the Bohr magneton is $\mu_B=9.2 \; \mbox{Gauss}\cdot\mbox{nm}^3$, and then we obtain the numerical value of $\varepsilon^2$:
\begin{equation}
\varepsilon^2=5.3\times 10^{-4} \; \mbox{nm}^{-3}.
\label{nvarep}
\end{equation}
Finally we obtain the free energy, which is the normal state gap:
\begin{equation}
F = 0.5 \; \mbox{meV.nm}^{-3}.
\label{gapm}
\end{equation}
For comparison we give the approximate values for the superconducting gap density of metals and cuprates. They are $10^{-4} \;\mbox{meV.nm}^{-3}$ and
$10 \;\mbox{meV.nm}^{-3}$, respectively. This follows from the BCS formula $F_{gap}=2\Delta n_{g}$, where $2\Delta$
is the energy required to break a single Cooper pair and $n_{g}$
represents the density of available Cooper pairs, namely
$n_g=0.187\Delta n/E_F$, $n$ and $E_F$ being the electronic density
and the Fermi energy, respectively. for metals $n \sim 0.1\,\mbox{nm}^{-3}$, $\Delta/E_F \sim 10^{-4}$ and
$2\Delta \sim 1.0 \,\mbox{meV}$, while for the cuprates
the gap is ten times larger than that of metals,
$2\Delta \sim 10 \,\mbox{meV}$, and $n_g\sim 1.0 \,\mbox{nm}^{-3}$,
since there are a few Cooper pairs~\cite{batlogg91} occupying the
coherence length volume, $\xi_{ab}^2\xi_c$, where $\xi_{ab}\sim 1.5
\,\mbox{nm}$, $\xi_c \sim 0.3 \, \mbox{nm}$.
In the next section we show the stability of this inhomogeneous state is due to topology, which prevents its decay into the lower energy homogeneous state ($F=0$).
\begin{table}[H]
	\centering
		\begin{tabular}{|c|c|c|c|}
		\hline
		\textbf{m} & \multicolumn{3}{c|}{$\Psi_{m}$}  \\ \cline{2-4}
		    & \textbf{Q} & $\frac{d}{L_{max}}$ & $J_3$ \\ \hline
		$-4$ & $+1$ & $0.01$ & $-\frac{7}{2}\hbar$  \\ \hline
		$-3$ & $+2$ & $0.01$ & $-\frac{5}{2}\hbar$  \\ \hline
		$-2$ & $-2$ & $0.01$ & $-\frac{3}{2}\hbar$  \\ \hline
		$-1$ & $-1$ & $0.01$ & $-\frac{1}{2}\hbar$  \\ \hline
		$0$  & $+1$  & $1.2$ & $+\frac{1}{2}\hbar$  \\ \hline
		$+1$ & $+2$ & $0.01$ & $+\frac{3}{2}\hbar$  \\ \hline
		$+2$ & $-2$ & $0.01$ & $+\frac{5}{2}\hbar$  \\ \hline
		$+3$ & $-1$ & $0.01$ & $+\frac{7}{2}\hbar$  \\ \hline
		\end{tabular}
	\caption{Limits of the skyrmionic charge for the eigenvectors $\Psi_{m,\pm}$}
	\label{tab1}
\end{table}

\section{The topological charge}
The order parameter $\Psi_m$ describes a lattice of skyrmions and to understand its topological properties we invoke the Moebius strip, formed  by rotating one end of 180$^\circ$ and joining it with the other end. Once glued in this way it cannot be returned to the 0$^\circ$ configuration. This introduces a discrete number which counts the number of twists present in the Moebius strip.
In case of the skyrmions the number of skyrmion cores within a layer, play the role of the twists in the Moebius strip.
Figs.(\ref{j3half}), (\ref{j3threehalf}), and (\ref{j3fivehalf}) show these skyrmion cores because superficial supercurrents circulate around them.
The skyrmion cores are sinkholes and at the same sources of magnetic field stream lines that cross the layer. These streamlines form closed loops around a given layer, which is then perforated twice. The condition that $\vec \nabla \cdot \vec h=0$ cannot be violated makes the rupture of these streamline loops nearly impossible, fact that brings stability to the $\Psi_m$ skyrmion solution.
Thus similarly to the twists in the Moebius strip, the closed loops of magnetic field streamlines cannot be made disappear in favor of a no loop configuration. From a mathematical perspective the topological stability stems from the existence of two closed surfaces that can be mapped one into the other, namely, a torus into a sphere. The torus is a consequence of the lattice pattern since the unit cell has periodic boundary conditions. The sphere is a consequence of the global rotational symmetry $SU(2)$ of the order parameter that can be identified with an $S^2-$sphere. The mapping of the torus into the sphere defines the second homotopy group of $S^2$, $\Pi_2(S^2)=Z$, which is the topological charge, or skyrmionic number. Thus the integer $Z$ defines the number of possible magnetic field configurations around a layer, which is ultimately defined by the number of cores. This is the  skyrmion's topological charge and is given by the following expression,
\begin{equation}
\textbf{Q}=\frac{1}{4\pi}\int d^2x \left[\left(\frac{\partial \hat{h}}{\partial x_1}\times\frac{\partial \hat{h}}{\partial x_2}\right).\hat{h}\right]_{x_3=0},
\label{skynumb}
\end{equation}
where $\hat{h}=\frac{\vec{h}}{|\vec{h}|}$. The topological charge is not invariant under the time reversal symmetry operation $\vec h \rightarrow - \vec h$. We find that this skyrmionic topological charge is a function of the angular momentum $m$, as shown in Table (\ref{tab1}). Table \ref{tab1} becomes cyclic for $m>3$, which means the state $m=0$ has the same topological charge of the $m=4$ state, the state $m=1$ has the same charge of the $m=5$ state and so on. To calculate the topological charge $\textbf{Q}$ we integrate numerically Eq.(\ref{skynumb}) using the software Mathematica.
We find an upper bound limit to the existence of the topological with regard to the unit cell size, called $L_{max}$, but not a lower bound limit. The checkerboard pattern $L=1.6\; nm$ falls within this limit, $L<L_{max}$. Surprisingly $L_{max}$ for the case  $m=0$ is different from the remaining ones, and there are skyrmions only in the range $1.2\lesssim\frac{d}{L}< \infty$, whereas for the remaining it holds that $0.01<\frac{d}{L}<\infty$.

A remarkable property of the present model is that the mean magnetic field value taken in any two-dimensional unit cell parallel to the layer at $0<x_3<d$ is
 \begin{equation}
\int_{x_3} \frac{d^2x}{A}\vec{h}=-16\pi\mu_{\beta}\varepsilon^2\left[\frac{1}{\sinh^2\left(\frac{\pi d}{L}\right)}+\frac{1}{\sinh^2\left(\frac{\sqrt{2}\pi d}{L}\right)}\right]\hat x_3.
\label{ulimit3}
\end{equation}
exactly as given in Eq.(\ref{ulimit}).

Further understanding of the present model is brought by Figs.(\ref{j3half}), (\ref{j3threehalf}) and (\ref{j3fivehalf}). They contain plots that are all obtained from the order parameter $\Psi_m$, given by Eq.(\ref{psipm}).
The $m$ states are also called by their $J_3$ values, namely, $\hbar(m+\frac{1}{2})$, according to Eq.(\ref{j3e}). The cases $J_3=\pm \frac{1}{2}\hbar$ ($m=0 \,\mbox{and} \,m=-1$), $\pm \frac{3}{2}\hbar$ ($m=1 \,\mbox{and} \,m=-2$), and $\pm \frac{5}{2}\hbar$ ($m=2 \,\mbox{and} \,m=-3$) are shown in Figs.(\ref{j3half}), (\ref{j3threehalf}) and (\ref{j3fivehalf}), respectively. Each of these figures contains four plots associated to the same $J_3$ modulus,
two for the positive $J_3$ state, and the other two for the negative $J_3$  state. The surface supercurrent $\vec J_s$ and the local magnetic field component $h_3$ are shown along a selected  layer.
The surface supercurrent is shown through arrows that indicate its circulation within the layer and the $h_3$ three-dimensional plot has the interesting feature of having  positive and negative values, confirming the presence of magnetic filed stream lines crossing the layer.
Thus the subplots of Figs.(\ref{j3half}), (\ref{j3threehalf}) and (\ref{j3fivehalf}) correspond to the following. Subplots (a) and (b) display $h_3$ and $J_s$ for the $m$ state, respectively, whereas subplots (c) and (d) are associated to the $h_3$ and $J_s$  $-m-1$ state, respectively. Comparison between subplots (b) and (d) for each of these three figures show that the surface supercurrents for the cases $m$ and $-m-1$ are the reverse of each other. Comparison between subplots (a) and (c) indicate an apparent reversal of $h_3$ sign for all the three figures. In fact this reversal does not exist because subplots (a) and (c) must have the same mean value given by Eq.(\ref{cms3}). They do have the same  mean magnetic field sign (negative) determined by Eq.(\ref{ulimit}).
Therefore it is quite remarkable that
Figs.(\ref{j3half}), (\ref{j3threehalf}) and (\ref{j3fivehalf})
display subplots with both senses of angular momentum, which means surface supercurrents circulating in opposite senses and yet the mean magnetic field points in one single direction for all cases. We look in more details each one of the figures.
Subplots (a) and (b) of Fig.(\ref{j3half}) describe the $J_3=\frac{1}{2}\hbar$ $(m=0)$ state. Indeed periodic boundary conditions show that each corner carry one fourth of the circulation of a single skyrmion at the unit cell in agreement with Table \ref{tab1} that gives, for this case, that $Q=+1$. The corner of the unit cell has $h_3<0$ and is where the skyrmion core is located.
Subplots (c) and (d) of Fig.(\ref{j3half}) describe the $J_3=-\frac{1}{2}\hbar$ state, whose supercurrent circulation is just in the opposite direction, and $Q=-1$. The $h_3$ plot is not the reverse of the previous case, as it seems to be, since both have the same value of $\int \frac{d^2x}{A} h_3$. Subplots (a) and (b) of Fig.(\ref{j3threehalf}) describe the $J_3=\frac{3}{2}\hbar$ $(m=1)$ state. Nevertheless in this case near to the corner there two skyrmions instead, and according to Table \ref{tab1}  $Q=+2$. The corner of the unit cell has $h_3<0$ and is where the skyrmion core is located.
Subplots (c) and (d) of Fig.(\ref{j3half}) describe the $J_3=-\frac{1}{2}\hbar$ state, whose supercurrent circulation is just in the opposite direction and $Q=-2$. Fig.(\ref{j3fivehalf}) shows circulation around the center of the unit cell in the opposite direction of previous figures.
Subplots (a) and (b) of Fig.(\ref{j3fivehalf}) describe the $J_3=\frac{5}{2}\hbar$ $(m=2)$ state. According to Table \ref{tab1}  $Q=-2$. Subplots (c) and (d) of Fig.(\ref{j3half}) describe the $J_3=-\frac{1}{2}\hbar$ state, whose supercurrent circulation is just in the opposite direction and $Q=+2$.

\section{Conclusions}
We show here that a very simple theoretical framework upholds properties to describe a topological state, that we conjecture to be the pseudogap. To account for this the theory must contain at least two order parameters.
We have studied the angular momentum properties of this state and calculated the gap that separates it from the homogeneous state under the condition that the local magnetic field falls below the threshold set by NMR/NQR and  $\mu$SR experiments. This inhomogeneous state is a lattice of skyrmions that breaks time reversal symmetry and leads to the checkerboard pattern.

\begin{acknowledgements}
The authors thank Edinardo Ivison Batista Rodrigues for allowing the use of Figs.(\ref{f1}) and (\ref{f2}) from his master thesis entitled "The Lichnerowicz-Weitzenb\"ock formula applied to superconductors with one and two-components".  Mauro M. Doria and Alfredo. A. Vargas-Paredes acknowledge the Brazilian agency CNPq and Inmetro for financial support.
\end{acknowledgements}

\bibliography{bibliog}

\end{document}